\pdfoutput=1
\documentclass[conference]{IEEEtran}
\IEEEoverridecommandlockouts
\usepackage{cite}
\makeatletter
\let\IEEEorigmakecaption\@makecaption
\long\def\@makecaption#1#2{%
  \ifx\@captype\@IEEEtablestring
    \footnotesize\bgroup\par\centering\@IEEEtabletopskipstrut{\normalfont\footnotesize #1.}\nobreakspace\nobreakspace{\normalfont\footnotesize\scshape #2}\par\addvspace{0.5\baselineskip}\egroup
    \@IEEEtablecaptionsepspace
  \else
    \IEEEorigmakecaption{#1}{#2}%
  \fi}
\makeatother
\usepackage{amsmath,amssymb}
\usepackage{graphicx}
\usepackage{booktabs,tabularx}
\usepackage{xcolor}
\usepackage{url,pifont}

\title{\fontsize{22.5}{27}\selectfont TAMER: Taming Write Amplification and Wear to Extend High-Bandwidth Flash Lifetime for LLM Serving}

\makeatletter
\def\@IEEEauthorblockAstyle{\normalfont\small}%
\def\@IEEEauthorblockAinterlinespace{2.35ex}
\def\@IEEEauthorblockconfadjspace{-7.5pt}%
\makeatother
\IEEEaftertitletext{\vspace{-\baselineskip}}%
\author{\IEEEauthorblockN{Yi Sun\textsuperscript{1,\dag}, Weikai Xu\textsuperscript{1,\dag,*}, Shuzhang Zhong\textsuperscript{2}, Xin Fan\textsuperscript{1}, Qi Liu\textsuperscript{1}, Tongbin Zhao\textsuperscript{2}, Meng Li\textsuperscript{2}, Yuan Xie\textsuperscript{1}}
\IEEEauthorblockA{\textsuperscript{1}The Hong Kong University of Science and Technology, Hong Kong, China; \textsuperscript{2}Peking University, Beijing, China\\
\textsuperscript{\dag}Equal contribution. \textsuperscript{*}Corresponding author: eexuweikai@ust.hk}}

\begin{document}
\maketitle

\begin{abstract}
High-bandwidth flash (HBF) offers read bandwidth comparable to that of high-bandwidth memory at a much higher capacity, enough to hold the weights and the key-value (KV) cache of large language models (LLMs).
However, flash wears out after a limited number of program/erase cycles, and HBF is integrated into the package and cannot be replaced, so wear leveling is essential for HBF-based LLM serving systems.
Yet no existing HBF design provides it, and the device fails far earlier than its erase budget allows, because per-request KV eviction amplifies writes, block selection ignores erase counts, and the blocks that hold weights are never erased.
To address these challenges, we propose TAMER, the first runtime-level wear-leveling framework for HBF without hardware flash translation layer, which requires no hardware modification, removes almost all write amplification with session packing and garbage collection, and evens out wear with wear-aware block selection.
Moreover, TAMER introduces periodic weight-KV rotation to migrate static weights onto heavily worn blocks, allowing all blocks to evenly share the erase budget.
Evaluations across four LLMs and four production traces show that TAMER reduces write amplification to at most 1.01 and extends HBF lifetime by 8.5$\times$ to 123.1$\times$ with negligible overhead, demonstrating the strong potential of durable HBF-based next-generation AI infrastructure.
\end{abstract}

\section{Introduction}
Large language models (LLMs) continue to grow in both parameter count and context length~\cite{mapatterson,deepseekv4}, and the key-value (KV) cache that avoids repeated prefill computation grows with them~\cite{vaswani2017,kwon2023vllm,gao2024cachedattention,liu2024cachegen,qin2025mooncake}.
High-bandwidth memory (HBM) alone has too little capacity and too high a cost to hold all of this state for long-context and high-concurrency inference.
High-bandwidth flash (HBF) stacks NAND dies behind a wide interface and operates thousands of planes in parallel, providing read bandwidth comparable to that of HBM at a much higher capacity density~\cite{ilkbahar,h3,flashaccel,hbflex,tilelens,memexplorer,haven}.
To break the HBM capacity wall, a full-HBF system dedicates the entire memory budget to flash and serves both the static model weights and the dynamic KV cache from it~\cite{hbflex}.
Unlike HBM, however, flash wears out, because each block endures only a limited number of program/erase cycles.
HBF also sits next to the processor in an advanced package and cannot be replaced once worn out~\cite{son}, so its endurance bounds the service life of the whole package and makes wear leveling essential.

\begin{figure}[!t]
\centering
\includegraphics[width=\columnwidth]{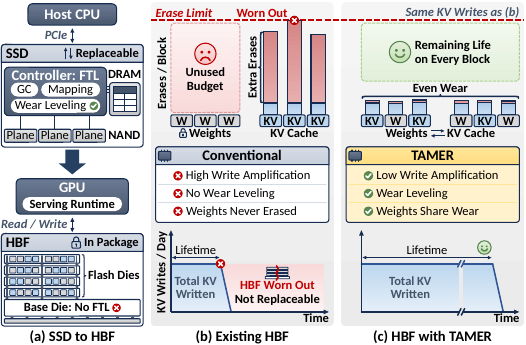}
\caption{Wear management in (a) an SSD, (b) existing HBF designs, and (c) TAMER. Existing designs lack wear leveling, and the flash cannot be replaced. TAMER lowers write amplification and levels wear in the serving runtime.}
\label{fig:budget}
\vspace{-10pt}
\end{figure}

Because the KV cache is rewritten continuously as requests arrive and are evicted, recent work reduces its flash writes.
FlashAccel dedicates blocks to each request~\cite{flashaccel}, HBFlex packs KV pages by lifetime and defers reclamation~\cite{hbflex}, and other studies bound the write volume through placement or scheduling~\cite{kyung,splash,hotcold,charhbf}.
These designs estimate the lifetime from the total write volume and the rated cycle count.
Solid-state drives (SSDs) face the same limit, but their controllers level wear across blocks behind a flash translation layer (FTL)~\cite{agrawal2008,gupta2009dftl}, a technique studied for two decades~\cite{chang2007static,chang2007dualpool,jung2007group,murugan2011rejuvenator}, and a worn-out drive can simply be replaced~\cite{hbfsucks}, as Fig.~\ref{fig:budget}(a) shows.

However, as Fig.~\ref{fig:budget}(b) shows, existing HBF designs do not level wear, and their lifetime estimates assume that all blocks wear equally.
Conventional SSD wear leveling does not carry over: to maximize bandwidth and avoid translation overhead, HBF omits a hardware FTL and exposes physical blocks directly to the serving runtime, where read bandwidth also depends on plane-level data striping.
Conversely, the serving runtime has direct visibility into request lifecycles and static weights that hardware controllers lack, making runtime-level wear management essential to address three wear challenges in full-HBF systems.
\textbf{\textit{Challenge~1:}} KV entries are allocated and evicted per request or multi-turn session, whereas flash is erased per block, so each erase copies valid pages of co-located requests and amplifies writes.
\textbf{\textit{Challenge~2:}} Allocation and reclamation ignore erase counts, so erases spread unevenly across the KV blocks.
\textbf{\textit{Challenge~3:}} The blocks that hold weights are never erased, so all wear falls on the KV blocks.
The most worn block thus fails while part of the erase budget remains unused.
Together, these factors severely limit the operational lifetime of HBF-based LLM serving systems.

This paper presents \textbf{TAMER}, the first wear-leveling design for HBF, which extends HBF lifetime in LLM serving by \underline{t}aming write \underline{am}plification and w\underline{e}a\underline{r}.
It erases a block only when the block holds nothing worth keeping, directs erases to the least worn blocks, and keeps long-lived data on the most worn ones.
As Fig.~\ref{fig:budget}(c) shows, TAMER maintains the address tables and the wear metadata in the serving runtime with negligible overhead, requires no change to the HBF device, and addresses each challenge with one mechanism.
Our contributions are as follows:
\begin{itemize}
\item \textbf{Session packing and garbage collection:} To reduce the write amplification of per-request eviction, we keep the KV pages of a session in the same blocks and defer reclamation until blocks turn entirely invalid, so that almost no valid pages are copied before an erase.
\item \textbf{Wear-aware block selection:} To even out wear across KV blocks, we allocate the least erased free block and reclaim the least erased invalid blocks first within each plane, so no block is favored by its position in a free list.
\item \textbf{Rotation between weight and KV blocks:} To let the blocks that hold weights share the wear, we periodically move the weights onto heavily erased blocks and return the lightly erased blocks they vacate to the KV cache.
\item We evaluate TAMER on all 16 combinations of four models and four production traces and find that it lowers write amplification to at most 1.01 and extends HBF lifetime by 8.5$\times$ to 123.1$\times$, while rotation adds at most 0.6\% erases and the new erase counts and weight locations take at most 9\,MB of host memory for each 3\,TiB device.
\end{itemize}
\section{Background and Motivation}
\subsection{KV Write Characteristics of LLM Serving}
\begin{figure}[!t]
\centering
\includegraphics[width=\columnwidth]{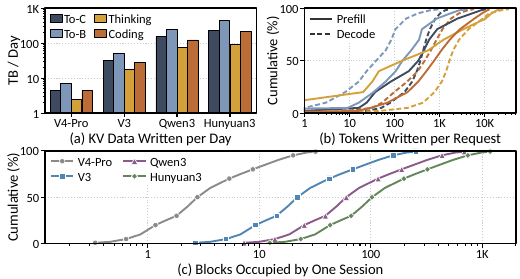}
\par\vspace{-5pt}%
\caption{KV writes in LLM serving, showing (a) the daily volume per device, (b) the tokens per request, and (c) the blocks per session on the To-C trace.}
\label{fig:workload}
\end{figure}

We study four models~\cite{deepseekv4,deepseekv3,qwen3,hy3}, listed in Table~\ref{tab:models}, on the To-C, To-B, Thinking, and Coding traces of a production LLM service~\cite{kvcachewild}, which record consumer chat, business API, reasoning, and coding requests.
The KV data that one device writes per day, shown in Fig.~\ref{fig:workload}(a), scales with the KV size per token and spans two orders of magnitude across the models.
A request writes the KV entries of its new prompt tokens in one burst during prefill and then appends the entries of each generated token during decode.
The amount written per request varies by more than two orders of magnitude within a trace, as Fig.~\ref{fig:workload}(b) shows, and prefill dominates the written data on all traces but Thinking.

The KV state of a session is also large relative to a flash block.
For three of the four models, almost every session spans multiple blocks and half of the sessions span more than twenty, as Fig.~\ref{fig:workload}(c) shows.
Blocks that hold a single session become invalid together when the session is evicted.
Blocks filled in arrival order instead mix the pages of many requests~\cite{hbflex}, and their pages become invalid one request at a time.

\begin{table}[!htb]
\vspace{-6pt}%
\caption{Evaluated models with KV and weight footprints.}
\label{tab:models}
\centering
\footnotesize
\begin{tabular}{@{}lrrrr@{}}
\toprule
Model & Layers & KV bytes/token & Devices & Weight blocks \\
\midrule
DeepSeek-V4-Pro & 61 & 4,460 & 2$^{*}$ & 24\% \\
DeepSeek-V3 & 61 & 35,136 & 1 & 20\% \\
Qwen3-235B & 94 & 96,256 & 1 & 7\% \\
Hunyuan3 (Hy3) & 80 & 163,840 & 1 & 9\% \\
\bottomrule
\end{tabular}\\[2pt]
{\scriptsize\raggedright The last column gives the share of the blocks of the system that hold weights.\\ $^{*}$Its weights alone would fill 48\% of the blocks of one device.\par}
\vspace{-8pt}%
\end{table}

\subsection{High-Bandwidth Flash and Its Endurance}
\begin{figure}[!t]
\centering
\includegraphics[width=\columnwidth]{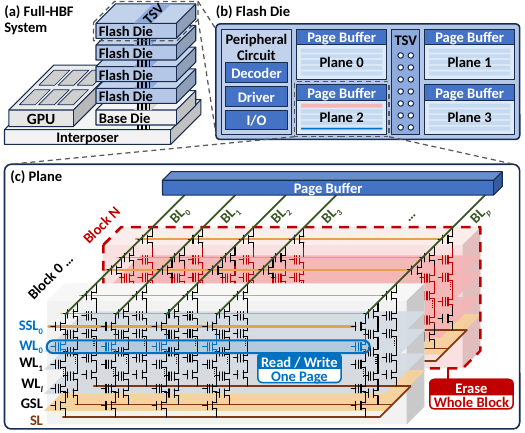}
\caption{Organization of an HBF stack into dies, planes, blocks, and pages. Pages are read and programmed individually, and a block is erased as a whole.}
\label{fig:hbf}
\vspace{-10pt}
\end{figure}

Fig.~\ref{fig:hbf} illustrates how HBF is organized into stacks, dies, planes, blocks, and pages~\cite{ocphbf}.
Reads and programs operate on pages, but an erase clears a whole block, and a page cannot be reprogrammed until its block is erased.
Because HBF obtains its bandwidth by operating many planes concurrently, data must be spread across planes to be read at full speed.
Hybrid designs keep HBM alongside HBF~\cite{h3,flashaccel,memvillage,hbmhbfpool,atassi,yin}.
This paper targets full-HBF designs, which use only flash~\cite{hbflex}.

Each block endures a limited number of program/erase cycles~\cite{cai2017}.
Existing studies estimate the lifetime as
\begin{equation}
T_{\mathrm{life}} = \frac{C \cdot N_{\mathrm{PE}}}{R_W \cdot \mathit{WAF}},
\label{eq:life}
\end{equation}
where $C$ is the flash capacity, $N_{\mathrm{PE}}$ is the rated cycle count, $R_W$ is the write rate, and $\mathit{WAF}$ is the write amplification factor~\cite{kyung,hbflex,splash,hotcold,charhbf,dash,lachesis}.
Equation~\eqref{eq:life} holds only if all blocks are erased equally often.
In practice, the blocks that wear out first set the device lifetime.
As Table~\ref{tab:prior} shows, existing HBF studies avoid or reduce write amplification or assume that it is small, yet all of them assume uniform wear and none levels it.

SSDs level wear dynamically, by allocating less worn blocks to new writes, and statically, by relocating cold data so that lightly erased blocks return to use~\cite{chang2007static,chang2007dualpool,jung2007group,murugan2011rejuvenator}.
Both techniques run inside the drive controller, which sees only logical writes.
The serving runtime of HBF, in contrast, knows which request owns each page and which request will be evicted next.

\subsection{Wear Challenges in Full-HBF Systems}
\begin{figure}[!t]
\centering
\includegraphics[width=\columnwidth]{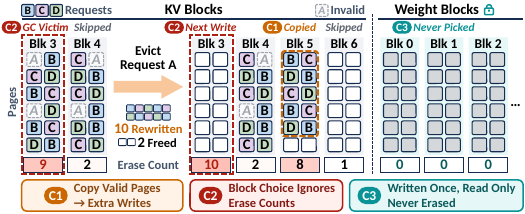}
\caption{Three wear challenges in a full-HBF system, marked C1 to C3, which are write amplification from per-request eviction, uneven wear across KV blocks, and weight blocks that do not share the wear.}
\label{fig:challenges}
\vspace{-5pt}%
\end{figure}

\begin{figure}[!t]
\centering
\includegraphics[width=\columnwidth]{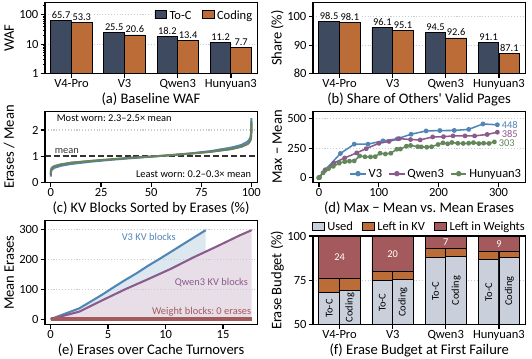}%
\caption{Wear under a basic KV manager, showing write amplification in (a)--(b), uneven wear across KV blocks in (c)--(d), and idle weight blocks in (e)--(f). Panels (c)--(e) use the To-C trace.}
\label{fig:evidence}
\vspace{-9pt}
\end{figure}

Fig.~\ref{fig:challenges} follows blocks through the erase loop of a full-HBF system and marks the three points, C1 to C3, at which lifetime is lost.
Fig.~\ref{fig:evidence} quantifies each of them under a basic KV manager without wear leveling, which is also the baseline in Section~IV-B.
\begin{table}[!t]
\caption{Write amplification and wear in HBF studies.}
\label{tab:prior}
\centering
\footnotesize
\begin{tabular}{@{}llll@{}}
\toprule
Study & WAF & Wear across blocks & Wear leveling \\
\midrule
H3~\cite{h3} & read-only HBF & -- & -- \\
FlashAccel~\cite{flashaccel} & avoided & assumed uniform & no \\
HBFlex~\cite{hbflex} & 1.04, measured & assumed uniform & no \\
SPLASH~\cite{splash} & 1.15, assumed & assumed uniform & no \\
\cite{kyung,hotcold,lachesis} & 1.02, assumed & assumed uniform & no \\
\cite{charhbf,dash,hbfsucks} & 1, assumed & assumed uniform & no \\
\midrule
\textbf{TAMER} & measured & counted per block & yes \\
\bottomrule
\end{tabular}
\vspace{-5pt}%
\end{table}

\textbf{Challenge 1: Write amplification from per-request eviction.}
KV entries are evicted per request, but flash is erased per block.
Many requests write at the same time, and a basic manager fills blocks in arrival order, which scatters the pages of one request over blocks that also hold pages of others.
Before such a block is erased, the live pages of other requests must be copied elsewhere, and every copy consumes cycles without storing new data~\cite{hu2009waf}.
On the To-C and Coding traces, these live pages fill 87\% to 98\% of the blocks of an evicted request, and the WAF reaches 7.7 to 65.7, as Fig.~\ref{fig:evidence}(a)--(b) show.
Private blocks per request avoid the copies but strand free pages in partly filled blocks, so the manager must control which pages share a block and when a block is reclaimed.

\textbf{Challenge 2: Uneven wear across KV blocks.}
Write amplification sets how many erases occur, and two choices decide which blocks absorb them, the free block that the next write takes and the block that reclamation erases.
A basic manager makes both without regard to erase counts, taking free blocks in queue order and erasing whatever blocks an eviction touches.
Some blocks are then erased more often than others by chance, and nothing corrects the difference.
The gap keeps growing, as Fig.~\ref{fig:evidence}(c)--(d) show, and the device fails with its most worn block while the others still have cycles left.
Leveling this wear requires erase counts to be considered whenever a block is allocated or reclaimed.

\textbf{Challenge 3: Weight blocks do not share the wear.}
Weights are written once and read for every token, so their blocks are never erased, and the KV blocks, which are only part of the device, absorb all of the wear, as Fig.~\ref{fig:evidence}(e) shows.
When the first block fails, 7\% to 24\% of the erase budget of the system is still unused in the weight blocks, as Fig.~\ref{fig:evidence}(f) shows.
This endurance can be used only if weights move onto worn blocks, and because each move copies a whole block and costs erases, the moves must be rare and well targeted.

\section{TAMER Design}
\begin{figure*}[!t]
\centering
\includegraphics[width=\textwidth]{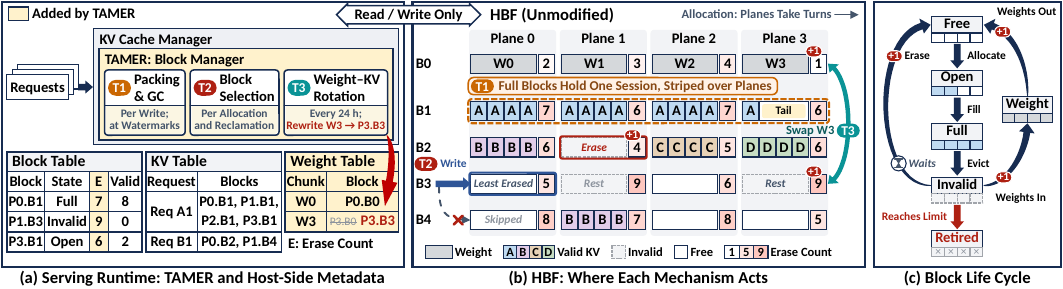}
\vspace{-20pt}
\caption{Overview of TAMER. (a) The serving runtime only reads and writes HBF, keeps the state, erase count, and valid pages of every block, and rewrites the physical locations of KV entries and weights when data move. (b) Where T1--T3 act. Full blocks hold one session and are striped over the planes, writes and erases go to the least erased blocks while the most erased rest, and weights swap places with resting blocks. (c) Life cycle of a block.}
\label{fig:arch}
\vspace{-15pt}
\end{figure*}

TAMER pursues the two goals of wear leveling.
It lowers write amplification so that fewer erases occur, through session packing and garbage collection, and it spreads the remaining erases evenly, through wear-aware block selection across the KV blocks and rotation between weight and KV blocks.
These three mechanisms, marked T1 to T3 in Fig.~\ref{fig:arch}, address the three challenges of Section~II-C in turn.

\subsection{System Architecture and Host-Side Metadata}
HBF exposes physical blocks without a translation layer, so the KV cache manager of the serving runtime already decides where every page is written.
TAMER extends this manager, as Fig.~\ref{fig:arch}(a) shows, and issues only reads and writes, because HBF erases a block implicitly when its first page is written~\cite{ocphbf}.
It keeps three tables.
The block table records the state, erase count, and valid pages of every block, the KV table records the blocks of every request, and the weight table records the block of every weight chunk.
Because the KV and weight tables hold physical locations, TAMER rewrites the affected entries whenever garbage collection or rotation moves data, and the old copy stays readable until its block is reused.
Most of this metadata exists in any KV cache manager.
TAMER adds the erase counts, the weight table, and a pointer to the partly filled tail block of each session.

Decisions are split between the device and its planes, as Fig.~\ref{fig:arch}(b) shows.
Device-wide policies determine which request is evicted, when reclamation and rotation start, and which plane the next allocation visits.
The block itself is always chosen within a plane, from heaps that order the free and the invalid blocks of that plane by erase count.
Data thus never leave their plane, and the planes that serve a session, and with them its read bandwidth, stay fixed.

Fig.~\ref{fig:arch}(c) shows the life cycle of a block.
An allocation opens a free block, which fills up and turns invalid once all of its pages are evicted.
Reclamation later returns it to the free pool and increments its erase count, and a block that reaches the erase limit is retired.
Within this cycle, session packing and garbage collection govern when blocks turn invalid and are reclaimed, block selection governs which blocks are taken, and rotation decides which blocks leave to hold weights.

\subsection{Session Packing and Garbage Collection}
An erase causes no write amplification when every page of its block is already invalid.
To approach this case, session packing keeps pages that will be evicted together in the same blocks, and garbage collection reclaims a block only after all of its pages have been evicted.

\begin{figure}[!t]
\centering
\includegraphics[width=\columnwidth]{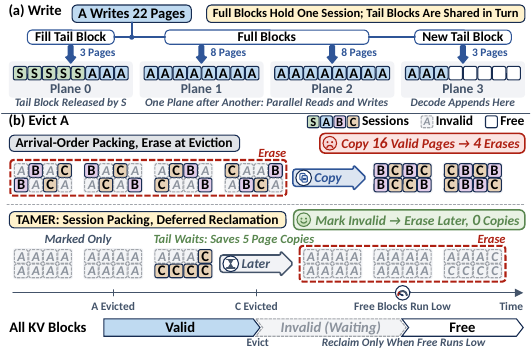}
\vspace{-20pt}
\caption{Session packing and garbage collection for a session A. (a) A fills a released tail block, two full blocks, and a new tail block, and (b) evicting A copies 16 pages with arrival-order packing but none with TAMER.}
\label{fig:packing}
\vspace{-15pt}
\end{figure}

Because the turns of a session extend one another and are evicted together, each session receives blocks of its own.
As Fig.~\ref{fig:packing}(a) shows, a write first fills the tail block of its session or, for the first write of a request whose session has none, the oldest tail block released by another session.
It then writes whole blocks, one from each plane in turn, so that they are written and read in parallel.
The remaining pages go to a new tail block, to which decode appends.

A session releases its tail block once it has no request in flight, since tail blocks kept private would strand their free pages.
A shared tail block therefore holds only sessions that were active at nearly the same time, and such sessions tend to be evicted close together.

Garbage collection keeps eviction and erase apart.
When valid pages reach 90\% of the KV capacity, TAMER evicts the least recently used completed requests until 85\% remains.
A request that extends an evicted one can no longer be reused and is evicted with it.
Eviction merely marks pages invalid.
Reclamation waits until 95\% of the KV blocks have been written, and each plane then reclaims entirely invalid blocks until 10\% of its KV blocks are free.
This delay gives shared tail blocks time to turn entirely invalid, so garbage collection rarely copies a page.
In Fig.~\ref{fig:packing}(b), evicting A copies 16 pages under arrival-order packing but none under TAMER.

Valid pages are copied only when a plane has no entirely invalid block left.
The plane then reclaims a block that is at most half valid, fewest valid pages first, after copying its valid pages to a relocation block in the same plane.
A plane down to its last free block starts reclaiming on its own, and if no plane has a free block, TAMER evicts about 1\% of the KV capacity at a time until a block is freed.

\subsection{Wear-Aware Block Selection}
Block selection spreads the erases evenly across the KV blocks.
A block is chosen at two points, when a free block is allocated and when an invalid block is reclaimed, and TAMER consults erase counts at both.

Allocation follows a cursor that cycles through the planes and takes the least erased free block of the current plane.
Since the choice depends only on erase counts, no block is favored or starved by its position in a free list, and a worn block is taken only after the less worn blocks of its plane have caught up.
Allocation has the larger effect of the two, because every erased block re-enters use through it.

\begin{figure}[!t]
\centering
\includegraphics[width=\columnwidth]{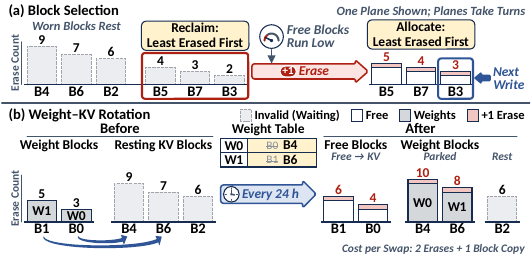}
\caption{Wear-aware block selection and rotation. (a) The less erased invalid blocks are reclaimed first, and writes take the least erased free block. (b) The least erased weight blocks move onto the most erased resting blocks.}
\label{fig:select}
\vspace{-10pt}
\end{figure}

Reclamation likewise takes the least erased of the entirely invalid blocks first, and it stops as soon as enough blocks are free, which leaves the most erased blocks invalid and at rest.
When valid pages must be copied, candidates are ranked by valid pages, in steps of 5\% of a block, and then by erase count, both in increasing order.
Fig.~\ref{fig:select}(a) illustrates both rules.

\subsection{Rotation Between Weight and KV Blocks}
Rotation lets the blocks that hold weights share the wear.
It periodically moves weights onto worn blocks and hands the lightly erased blocks that they vacate to the KV cache.
Every 24 hours, at low load, each plane takes its four least erased weight blocks as sources.
The candidate targets are free blocks, entirely invalid blocks, and idle blocks with at most 5\% valid pages.
They are ranked by the erase count that they will have once they hold weights, which includes one more erase for targets that are not yet free.
Sources and targets are then paired, the least erased source with the most erased target.
A pair is swapped only if the target still ends up more erased than the source after counting the erase that returns the source to the KV cache, since the swap would otherwise spend an erase without evening out wear.

To swap, TAMER copies any valid pages of the target within the plane, writes the weights into the target, updates the weight table, and returns the source to the free pool, where block selection soon reuses it.
Each swap costs at most two erases and about one block copy, and a round rewrites at most 24\,GiB of weights across the device, about 0.17\,s of programs and erases per plane.
Rotation needs no threshold on the wear gap, because reclamation leaves the most erased blocks resting and these are the blocks that rotation targets.
The 24-hour period balances the lag of resting weight blocks against the erases that rotation adds and keeps both below 1\% of the lifetime.
Fig.~\ref{fig:select}(b) shows one round that moves two weight blocks.

\begin{figure*}[!t]
\centering
\includegraphics[width=\linewidth]{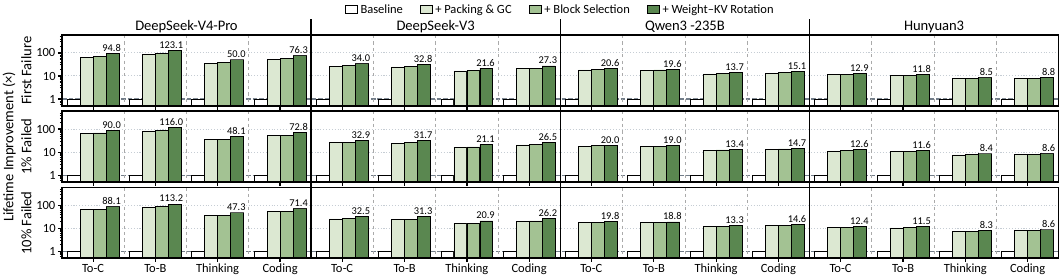}%
\vspace{-10pt}
\caption{Lifetime improvement over the baseline until the first block, 1\%, and 10\% of the blocks fail, as the three mechanisms are added in turn.}
\label{fig:lifetime}
\vspace{-15pt}
\end{figure*}
\section{Evaluation Methodology}
\subsection{Workloads and System Configuration}
We evaluate TAMER with a custom trace-driven simulator that replays the traces of Section~II-A and tracks the valid pages and the erase count of every block.
It models an H200-class GPU whose memory is the full-HBF device of Table~\ref{tab:config}, six stacks of 16 dies with 32 planes per die.
Requests arrive at the rate that the GPU sustains under full load, which sets the KV write rate.
DeepSeek-V4-Pro runs on two GPUs with tensor parallelism, each holding half of the weights and a full copy of the KV cache, so we simulate one of them.
\begin{table}[!htb]
\vspace{-5pt}%
\caption{System configuration.}
\label{tab:config}
\centering
\footnotesize
\setlength{\tabcolsep}{4pt}
\begin{tabularx}{\columnwidth}{@{}llX@{}}
\toprule
Component & Parameter & Configuration \\
\midrule
HBF & Planes $\times$ blocks $\times$ pages & 3,072 $\times$ 512 $\times$ 512 \\
 & Page size / capacity & 4\,KiB / 3\,TiB \\
 & Endurance / program / erase & 100,000 cycles / 75\,$\mu$s / 1.5\,ms \\
Eviction & Start / stop threshold & 90\% / 85\% of KV pages valid \\
Reclamation & Start threshold & 95\% of KV blocks written \\
 & Target free blocks & 10\% per plane \\
 & Maximum valid pages & 50\% per block \\
Rotation & Interval / weight blocks & 24\,h / 4 per plane \\
\bottomrule
\end{tabularx}
\vspace{-6pt}%
\end{table}

The lifetime is the time until the first block reaches 100,000 erases, and we also report when 1\% and 10\% of the blocks do.
Because no replay runs that long, we extrapolate these times from the erase counts measured in the replays.

\subsection{Baseline Configurations}
The baseline is the basic KV manager of Section~II-C.
It fills the open block of each plane in arrival order, takes free blocks from a per-plane FIFO queue, erases the blocks of each evicted request after copying out the valid pages of other requests, and never moves weights.
We then add the three mechanisms of TAMER one at a time, in the order of Section~III, and also evaluate the baseline with wear-aware block selection alone.
Because all configurations serve the same requests with the same eviction thresholds and their KV write rates differ by less than 1\%, the gains do not come from writing less.

\section{Evaluation}
\subsection{End-to-End Lifetime}
TAMER extends the lifetime by 8.5$\times$ to 123.1$\times$ over the baseline, as Fig.~\ref{fig:lifetime} shows, and the time until 1\% or 10\% of the blocks fail by 8.3$\times$ to 116.0$\times$.
Most of this gain, 7.5$\times$ to 83.3$\times$, comes from session packing and garbage collection and matches the write amplification that they remove.
It is largest for DeepSeek-V4-Pro, where each baseline block mixes many small sessions.
Wear-aware block selection adds another 3\% to 12\%, and rotation a further 8\% to 31\%.

In absolute terms, TAMER raises the lifetime from \mbox{0.2--1.1} to \mbox{2.0--12} years for Qwen3-235B and Hunyuan3, from \mbox{0.5--2.4} to \mbox{18--52} years for DeepSeek-V3, and from \mbox{1.0--7.2} to more than 100 years for DeepSeek-V4-Pro.
Even the shortest lifetime grows from two months to two years.
The baseline wears out within a year in 9 of the 16 combinations, whereas TAMER lasts more than five years in 12 of them.
\subsection{Session Packing and Garbage Collection}
\begin{figure}[!t]
\centering
\includegraphics[width=\linewidth]{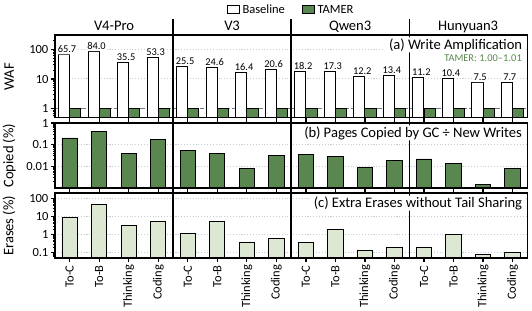}%
\vspace{-10pt}
\caption{Session packing and garbage collection, showing (a) the WAF, (b) the pages copied per newly written page, and (c) the erases added without shared tail blocks.}
\label{fig:evalpacking}
\vspace{-15pt}
\end{figure}

TAMER lowers the WAF from \mbox{7.5--84} to at most 1.01, as Fig.~\ref{fig:evalpacking}(a) shows.
In Fig.~\ref{fig:evalpacking}(b), the pages that garbage collection copies amount to at most 0.4\% of the newly written pages, so almost every block is entirely invalid when it is erased.
Fig.~\ref{fig:evalpacking}(c) isolates the effect of shared tail blocks.
Without them, the erases of DeepSeek-V4-Pro, whose sessions are small relative to a block, increase by 3.3\% to 49\%, and those of the other models by 0.1\% to 5.6\%.
On the To-B trace of DeepSeek-V4-Pro, almost all KV blocks are then occupied while only two thirds of their pages are valid, and a shortage of free blocks forces 96\% of the evictions.

\subsection{Block Selection and Wear Spread}
\begin{figure}[!t]
\centering
\includegraphics[width=\linewidth]{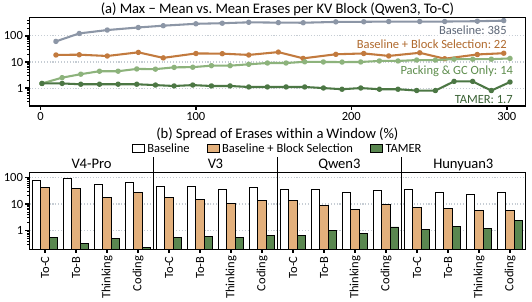}%
\vspace{-10pt}
\caption{Wear-aware block selection, showing (a) max-to-mean erase gap across KV blocks and (b) within-window erase spread.}
\label{fig:evalselect}
\vspace{-10pt}
\end{figure}

Fig.~\ref{fig:evalselect}(a) follows the gap between the highest and the mean erase count of the KV blocks for Qwen3-235B on To-C.
The gap grows to 385 under the baseline.
With session packing and garbage collection alone, it still drifts up to 14 because free blocks are reused in queue order.
Wear-aware block selection alone holds it at about 20, and TAMER keeps it below 2.
Fig.~\ref{fig:evalselect}(b) reports the coefficient of variation of erases across KV blocks over a window of about 90 erases per block.
It falls from \mbox{23--95\%} under the baseline to \mbox{6--41\%} with wear-aware block selection alone and to \mbox{0.2--2.4\%} with TAMER.

\subsection{Rotation and Overhead}
\begin{figure}[!t]
\centering
\includegraphics[width=\linewidth]{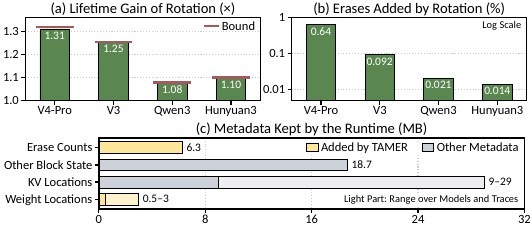}%
\vspace{-10pt}
\caption{Trace-averaged rotation results: (a) lifetime gain and bound, (b) added erases, and (c) runtime metadata (3\,TiB device).}
\label{fig:evalrotation}
\vspace{-9pt}
\end{figure}

Fig.~\ref{fig:evalrotation}(a) shows that rotation extends the lifetime by 1.08$\times$ to 1.31$\times$, more for models whose weights occupy more blocks, and that it reaches 97\% to 99\% of the gain that these blocks allow.
In Fig.~\ref{fig:evalrotation}(b), the erases that rotation adds amount to only 0.01\% to 0.6\%.
The erase counts and weight locations that TAMER adds take 7 to 9\,MB of the 35 to 57\,MB of host metadata for each 3\,TiB device, as Fig.~\ref{fig:evalrotation}(c) shows.

\section{Conclusion}
Write amplification and uneven wear, not the raw KV write volume, limit the lifetime of full-HBF systems.
TAMER, the first wear-leveling design for HBF, addresses both in the serving runtime.
Session packing and garbage collection remove almost all write amplification, wear-aware block selection evens out KV block wear, and rotation lets weight blocks share the erases.
Together, they lower the WAF to at most 1.01 and extend HBF lifetime by 8.5$\times$ to 123.1$\times$ across four models and four production traces.
TAMER achieves these gains with negligible extra erases and a tiny host memory footprint, indicating its great potential for durable LLM serving.

\clearpage\clearpage%
\bibliographystyle{IEEEtran}
\bibliography{refs}

\begin{thebibliography}{10}
\providecommand{\url}[1]{#1}
\csname url@samestyle\endcsname
\providecommand{\newblock}{\relax}
\providecommand{\bibinfo}[2]{#2}
\providecommand{\BIBentrySTDinterwordspacing}{\spaceskip=0pt\relax}
\providecommand{\BIBentryALTinterwordstretchfactor}{4}
\providecommand{\BIBentryALTinterwordspacing}{\spaceskip=\fontdimen2\font plus
\BIBentryALTinterwordstretchfactor\fontdimen3\font minus
  \fontdimen4\font\relax}
\providecommand{\BIBforeignlanguage}[2]{{%
\expandafter\ifx\csname l@#1\endcsname\relax
\typeout{** WARNING: IEEEtran.bst: No hyphenation pattern has been}%
\typeout{** loaded for the language `#1'. Using the pattern for}%
\typeout{** the default language instead.}%
\else
\language=\csname l@#1\endcsname
\fi
#2}}
\providecommand{\BIBdecl}{\relax}
\BIBdecl

\bibitem{mapatterson}
X.~Ma and D.~Patterson, ``Challenges and research directions for large language
  model inference hardware,'' \emph{Computer}, vol.~59, no.~5, pp. 55--64,
  2026.

\bibitem{deepseekv4}
{DeepSeek-AI}, ``{DeepSeek-V4}: Towards highly efficient million-token context
  intelligence,'' arXiv:2606.19348, 2026.

\bibitem{vaswani2017}
A.~Vaswani \emph{et~al.}, ``Attention is all you need,'' in \emph{Advances in
  Neural Information Processing Systems (NeurIPS)}, 2017.

\bibitem{kwon2023vllm}
W.~Kwon \emph{et~al.}, ``Efficient memory management for large language model
  serving with {PagedAttention},'' in \emph{Proc. ACM Symposium on Operating
  Systems Principles (SOSP)}, 2023, pp. 611--626.

\bibitem{gao2024cachedattention}
B.~Gao \emph{et~al.}, ``Cost-efficient large language model serving for
  multi-turn conversations with {CachedAttention},'' in \emph{Proc. USENIX
  Annual Technical Conference (ATC)}, 2024, pp. 111--126.

\bibitem{liu2024cachegen}
Y.~Liu \emph{et~al.}, ``{CacheGen}: {KV} cache compression and streaming for
  fast large language model serving,'' in \emph{Proc. ACM SIGCOMM Conference},
  2024, pp. 38--56.

\bibitem{qin2025mooncake}
R.~Qin \emph{et~al.}, ``Mooncake: Trading more storage for less computation --
  a {KVCache}-centric architecture for serving {LLM} chatbot,'' in \emph{Proc.
  USENIX Conference on File and Storage Technologies (FAST)}, 2025, pp.
  155--170.

\bibitem{ilkbahar}
A.~Ilkbahar, R.~Ramanujan, and T.~Coughlin, ``High bandwidth flash: A new
  memory for the next generation of {AI} computing,'' \emph{Computer}, vol.~59,
  no.~5, pp. 134--138, 2026.

\bibitem{h3}
M.~Ha, E.~Kim, and H.~Kim, ``{H3}: Hybrid architecture using high bandwidth
  memory and high bandwidth flash for cost-efficient {LLM} inference,''
  \emph{IEEE Computer Architecture Letters}, vol.~25, no.~1, pp. 49--52, 2026.

\bibitem{flashaccel}
X.~Wang \emph{et~al.}, ``{FlashAccel}: Leveraging high-bandwidth flash for
  high-throughput {LLM} inference,'' arXiv:2607.10186, 2026.

\bibitem{hbflex}
S.~Zhong \emph{et~al.}, ``{HBFlex}: A flexible memory system for bridging
  fine-grained {LLM} states and coarse-grained {HBF} parallel execution,''
  arXiv:2609.18675, 2026.

\bibitem{tilelens}
J.~H. Ju \emph{et~al.}, ``{TileLens}: Efficiently using large-granularity
  memory systems with transparent two-dimensional memory layout,''
  arXiv:2607.04031, 2026.

\bibitem{memexplorer}
H.~Wu \emph{et~al.}, ``{MemExplorer}: Navigating the heterogeneous memory
  design space for agentic inference {NPUs},'' arXiv:2604.16007, 2026.

\bibitem{haven}
P.-K. Hsu, W.~Xu, Q.~Liu, T.~Rosing, and S.~Yu, ``{HAVEN}: High-bandwidth flash
  augmented vector engine for large-scale approximate nearest-neighbor search
  acceleration,'' arXiv:2603.01175, 2026.

\bibitem{son}
D.~Son \emph{et~al.}, ``Exploring high-bandwidth flash for modern {LLM}
  inference: Opportunities and challenges,'' \emph{IEEE Computer Architecture
  Letters}, 2026.

\bibitem{kyung}
K.~Kyung, Y.~J. Moon, J.~Cho, and J.~H. Ahn, ``High-bandwidth flash for {KV}
  caches: Endurance and performance implications,'' \emph{IEEE Computer
  Architecture Letters}, vol.~25, no.~1, pp. 210--213, 2026.

\bibitem{splash}
A.~A. Jonnalagadda \emph{et~al.}, ``{SPLASH}: Co-designing sparse attention
  with high-bandwidth flash for efficient long-context inference,''
  arXiv:2609.23816, 2026.

\bibitem{hotcold}
J.~Baek, W.~Ji, S.~Yoo, and J.-Y. Kim, ``Hot--cold tiering of {HBM} and high
  bandwidth flash for agentic {LLM} serving,'' arXiv:2609.25782, 2026.

\bibitem{charhbf}
Z.~Yu \emph{et~al.}, ``Characterizing high bandwidth flash for {LLM} serving,''
  arXiv:2609.39131, 2026.

\bibitem{agrawal2008}
N.~Agrawal, V.~Prabhakaran, T.~Wobber, J.~D. Davis, M.~Manasse, and
  R.~Panigrahy, ``Design tradeoffs for {SSD} performance,'' in \emph{Proc.
  USENIX Annual Technical Conference (ATC)}, 2008.

\bibitem{gupta2009dftl}
A.~Gupta, Y.~Kim, and B.~Urgaonkar, ``{DFTL}: A flash translation layer
  employing demand-based selective caching of page-level address mappings,'' in
  \emph{Proc. International Conference on Architectural Support for Programming
  Languages and Operating Systems (ASPLOS)}, 2009, pp. 229--240.

\bibitem{chang2007static}
Y.-H. Chang, J.-W. Hsieh, and T.-W. Kuo, ``Endurance enhancement of
  flash-memory storage systems: An efficient static wear leveling design,'' in
  \emph{Proc. Design Automation Conference (DAC)}, 2007.

\bibitem{chang2007dualpool}
L.-P. Chang, ``On efficient wear leveling for large-scale flash-memory storage
  systems,'' in \emph{Proc. ACM Symposium on Applied Computing (SAC)}, 2007.

\bibitem{jung2007group}
D.~Jung, Y.-H. Chae, H.~Jo, J.-S. Kim, and J.~Lee, ``A group-based
  wear-leveling algorithm for large-capacity flash memory storage systems,'' in
  \emph{Proc. International Conference on Compilers, Architecture, and
  Synthesis for Embedded Systems (CASES)}, 2007.

\bibitem{murugan2011rejuvenator}
M.~Murugan and D.~H.~C. Du, ``Rejuvenator: A static wear leveling algorithm for
  {NAND} flash memory with minimized overhead,'' in \emph{Proc. IEEE Symposium
  on Mass Storage Systems and Technologies (MSST)}, 2011.

\bibitem{hbfsucks}
Z.~Li \emph{et~al.}, ``{HBF} sucks! {A} full-stack characterization of
  high-bandwidth flash for {KV}-centric {LLM} serving,'' arXiv:2608.11668,
  2026.

\bibitem{deepseekv3}
{DeepSeek-AI}, ``{DeepSeek-V3} technical report,'' arXiv:2412.19437, 2025.

\bibitem{qwen3}
A.~Yang \emph{et~al.}, ``{Qwen3} technical report,'' arXiv:2505.09388, 2025.

\bibitem{hy3}
{Tencent Hy Team}, ``{Hy3},'' Hugging Face model repository,
  https://huggingface.co/tencent/Hy3, 2026.

\bibitem{kvcachewild}
J.~Wang \emph{et~al.}, ``{KVCache} cache in the wild: Characterizing and
  optimizing {KVCache} cache at a large cloud provider,'' in \emph{Proc. USENIX
  Annual Technical Conference (ATC)}, 2025.

\bibitem{ocphbf}
{Open Compute Project}, ``{OCP HBF} architecture specification v0.7.0,''
  Technical Specification, 2026.

\bibitem{memvillage}
H.~Lee \emph{et~al.}, ``{Mem-Village}: A hybrid {HBF}-{HBM} architecture on
  glass for ultra-scale {AI} inference systems,'' in \emph{Proc. IEEE
  Electrical Design of Advanced Packaging and Systems (EDAPS)}, 2025.

\bibitem{hbmhbfpool}
J.~Park, H.~An, H.~Suh, Y.~Yoon, H.~Lee, and J.~Kim, ``{HBM}-{HBF}-centric
  memory pooling architecture with custom base die for terabyte-scale {LLM}
  inference,'' \emph{IEEE Computer Architecture Letters}, vol.~25, no.~2, pp.
  259--262, 2026.

\bibitem{atassi}
H.~Atassi, N.~Zilberman, and A.~Awad, ``Hardware-managed heterogeneous
  high-bandwidth memory and flash in {LLM} inference systems,'' \emph{IEEE
  Computer Architecture Letters}, vol.~25, no.~2, pp. 371--374, 2026.

\bibitem{yin}
Y.~Yin \emph{et~al.}, ``Potential applications of {HBF} in {LLM} serving
  systems,'' arXiv:2608.13127, 2026.

\bibitem{cai2017}
Y.~Cai, S.~Ghose, E.~F. Haratsch, Y.~Luo, and O.~Mutlu, ``Error
  characterization, mitigation, and recovery in flash-memory-based solid-state
  drives,'' \emph{Proceedings of the IEEE}, vol. 105, no.~9, pp. 1666--1704,
  2017.

\bibitem{dash}
S.~Kim, J.~Jin, and J.-Y. Kim, ``Beyond capacity: Scalable {MoE} {LLM}
  inference via high-bandwidth flash with direct {GPU} and {HBM} paths,''
  arXiv:2608.14333, 2026.

\bibitem{lachesis}
J.~Yang \emph{et~al.}, ``Lachesis: Lifetime-aware {KV} cache placement for
  agent serving across {HBM} and high-bandwidth flash,'' arXiv:2610.08378,
  2026.

\bibitem{hu2009waf}
X.-Y. Hu, E.~Eleftheriou, R.~Haas, I.~Iliadis, and R.~Pletka, ``Write
  amplification analysis in flash-based solid state drives,'' in \emph{Proc.
  Israeli Experimental Systems Conference (SYSTOR)}, 2009, pp. 1--9.

\end{thebibliography}
\end{document}